\documentclass[aps,pre,twocolumn,superscriptaddress,amsmath,amssymb,floatfix]{revtex4-2}
\usepackage{graphicx}
\usepackage{xcolor}
\usepackage{bm}
\usepackage{booktabs}
\usepackage[colorlinks=true,allcolors=blue]{hyperref}

\begin{document}

\title{On the richness of metamaterials generated by recursive cell division}

\author{Alexei Vazquez}
\email{alexei@nodeslinks.com}
\affiliation{Nodes \& Links Ltd, Salisbury House, Station Road, Cambridge, CB1 2LA, UK}

\date{\today}

\begin{abstract}
Scale-rich metamaterials are one member of a family of recursive cell divisions, which
differ in which cell is cut, where the cut passes and at what angle. We characterise six
members of this family through three network mappings, the cell tree, the dual graph of
ligaments and junctions, and the frame graph of spans, and through their elastic-wave
properties: the effective refractive index $n$, its isotropy and the band over which it
holds. The dual-graph degree law, the mean degree and the age law of the ligaments hit by
new ones are the same for every rule, so they cannot distinguish the materials. The cell
tree and the cell shapes do differ, and one consequence of them explains the physics: the
tail of long spans in the frame graph. With the longest spans as the length scale, the
bending part of the inverse squared index follows a single power law across rules,
densities and numbers of ligaments, and the same spans set the lowest local resonance and
the band edge. The stretching part, which caps the index, is carried by the straight frame
lines of the tiling rather than by the division. Random cut angles produce elongated cells and long spans, hence high index and
a narrow band; cuts perpendicular to the long axis of the cell keep the cells compact,
giving low index and a wide band. Cutting through the centroid instead of a random point
bounds the split fraction between $4/9$ and $5/9$, which removes the area hierarchy without
changing the index. The cut angle interpolates continuously between these limits, and the
isotropy peaks where the ligament orientations become isotropic while the cells are still
compact. In three dimensions the plate and strut divisions are stretching dominated and the
index stays low for every rule.
\end{abstract}

\maketitle

\section{Introduction}

Scale-rich (SR) metamaterials~\cite{both2026}, a recent example of irregular
metamaterial networks~\cite{wytock2026}, are grown one ligament at a time: a ligament
nucleates at a uniform random point and angle and grows until it meets the existing
structure, with thickness $\lambda_t=\lambda_0t^{-\alpha}$ for the $t$-th ligament.
Their unit cells reach effective refractive indices $1.5\le n\le8.3$, against
$1.0\le n\le1.9$ for square and hexagonal lattices, and an SR Luneburg lens focuses a
plane wave~\cite{both2026}. Geometrically the SR construction is Cowan's iterated cell
division: the cell holding the point, chosen in proportion to its area, is cut by a
chord~\cite{cowan2010,vazquez2026sr}. Its dual graph, with ligaments as nodes and
junctions as links, grows by a face-local rule and has a degree distribution with
exponent $\gamma=3$~\cite{vazquez2026sr}.

\begin{figure}[tb]
\includegraphics[width=\columnwidth]{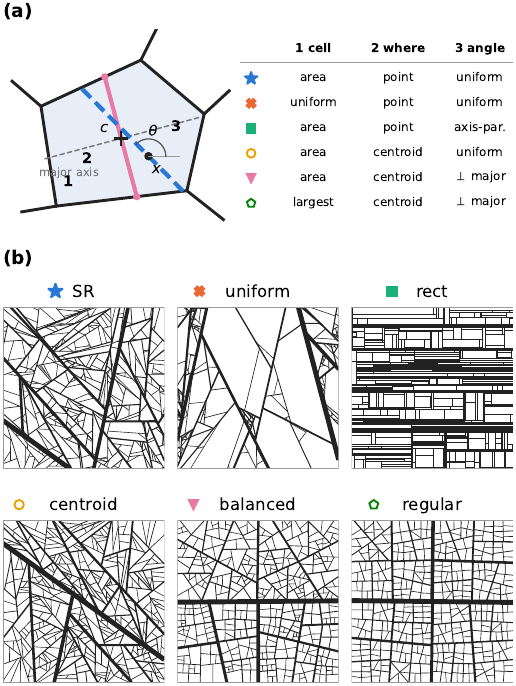}
\caption{\label{fig:rules}Recursive cell division. (a) One step: (1) a cell is picked (shaded),
(2) the cut passes through a uniform point $x$ or the centroid $c$, and (3) its angle $\theta$
is uniform, axis-parallel, or perpendicular to the major axis of the cell; the cut ends on
two sides of the cell, forming two T-junctions. Dashed blue: an SR cut (point $x$, random
$\theta$); solid pink: a \emph{balanced} cut (centroid, perpendicular to the major axis).
The table gives the three choices for each rule; every rule uses the thickness
$\lambda_t=\lambda_0t^{-\alpha}$. (b) The six rules, $T=400$ ligaments, line
width $\propto t^{-1/2}$ ($\alpha=1/2$). The symbol next to each name marks that structure
in every figure.}
\end{figure}

Cowan's construction defines a whole family of recursive divisions. A rule specifies
which cell is cut, where the cut passes, and at what angle. Changing these choices
produces materials that range from scale-rich hierarchies to nearly uniform cells
(Fig.~\ref{fig:rules}). All of them are generated just as simply, all accept the same
thickness hierarchy, and all are networks of straight ligaments meeting at T-junctions.
Each division can be read as several networks: the binary tree of cells, the dual graph
of ligaments, and the frame graph of spans between junctions, which is the elastic
structure itself.

In the long-wavelength regime a cellular solid of relative density $\bar\rho$ and
effective stiffness tensor $C$ carries quasi-longitudinal waves with velocity
$v(\hat{\bm k})$, the square root of the largest eigenvalue of the Christoffel matrix
$\Gamma_{ik}=C_{ijkl}\hat k_j\hat k_l$ divided by $\bar\rho$~\cite{auld1973}. The
effective refractive index $n(\hat{\bm k})=c_0/v(\hat{\bm k})$, with $c_0$ the velocity
in the constituent, is the design variable of gradient-index (GRIN) devices such as
elastic Luneburg lenses~\cite{tol2017,aguzzi2022}. A GRIN device realises a prescribed
profile $n(\bm r)$ with isotropic cells, and works up to the frequency at which the
homogenised index of its weakest cell fails. The figures of merit of a cell are
therefore its index, the isotropy of the index, and the band over which it holds. Cells
generated by different rules share the same square frame and can be tiled together; the
scattering at the interfaces between them is not studied here.

Here we ask a more general question: which structure among Cowan's recursive divisions
is best for elastic-wave control? For six division rules we characterise the network
mappings and the physical properties, and we identify which network properties explain
the differences between the physical properties. Balanced divisions, which cut each
cell through its centroid and perpendicular to its long axis, are best at low index, and
centroid divisions with random cut angles are best at high index.

Section~\ref{sec:div} defines the division rules, the three network mappings and the
mechanical model. Section~\ref{sec:laws} introduces the two relations that organise the
physics: the split of the index into a stretching ceiling and a bending term, and the
resonance of the weakest span. Section~\ref{sec:shared} shows which network properties are
the same for every rule. Section~\ref{sec:rules} characterises each of the six structures,
its network, its physics and the link between them. Section~\ref{sec:collapse} shows that
the long-span tail of the frame graph sets the index and the resonance of every rule, and
follows the cut angle continuously. Section~\ref{sec:best} ranks the structures at equal
index, Sec.~\ref{sec:3d} extends the comparison to three dimensions, and the Conclusions
summarise the results.

\section{Recursive divisions and their mappings}
\label{sec:div}

\subsection{Rules}
Starting from the unit square, step $t=1,\dots,T$ picks a cell and cuts it by a chord.
Cells are picked in proportion to area (SR), uniformly among the cells (\emph{uniform}),
or largest first (\emph{regular}). The cut passes through a uniform point of the cell
(SR, \emph{uniform}, \emph{rect}) or through its centroid (\emph{centroid},
\emph{balanced}, \emph{regular}). Its direction is uniform (SR, \emph{uniform},
\emph{centroid}), axis-parallel (\emph{rect}), or perpendicular to the major axis of the
second-moment tensor of the cell (\emph{balanced}, \emph{regular}). A one-parameter family
interpolates between \emph{balanced} and \emph{centroid}: the cut passes through the
centroid at an angle $\theta=\theta_\perp+\kappa\pi(u-\tfrac12)$, with $\theta_\perp$
perpendicular to the major axis and $u$ uniform in $(0,1)$; $\kappa=0$ is
\emph{balanced} and $\kappa=1$ is \emph{centroid}. Every rule uses the SR thickness
$\lambda_t=\lambda_0t^{-\alpha}$, with $\lambda_0$ set by $\bar\rho$.

Related constructions are well studied. STIT tessellations pick cells in proportion to
perimeter and cut them by isotropic random lines; they are iteration stable, and their
mean corner number, edge lengths and cell shapes are known
exactly~\cite{nagel2005}. Hierarchical crack patterns in drying layers divide domains one
at a time, across their long axis and close to their middle, as in the \emph{balanced}
and \emph{regular} rules, and produce mostly four-sided
domains~\cite{bohn2005,bohn2005b}. We use the rules above because they separate the three
choices: which cell, where, and at what angle.

\subsection{Mappings}
Each division defines three networks.
\emph{The cell tree} is the binary tree of cells, with the area fraction $f$ of the
smaller daughter at each split, the depth of each final cell and its area $A$. We also
record the corner number of each cell and its elongation
$e=\max_\theta w(\theta)/\min_\theta w(\theta)$, where $w(\theta)$ is the width of the cell
along the direction $\theta$, the extent of its projection (support-function width).
\emph{The dual graph} has the ligaments as nodes; a link joins the ligament born at a
T-junction (the stem) to the ligament it ends on (the host). We record its degree
distribution, clustering, assortativity and the number of fewest-turn paths between
ligaments, as in Ref.~\cite{vazquez2026sr}.
\emph{The frame graph} has the junctions as nodes and the spans between consecutive
junctions along a ligament as links. It is the elastic structure. We record the span
lengths $\ell$, the orientation fabric
$a_4=\sum\ell\cos4\phi/\sum\ell$, with $\phi$ the span orientation ($a_4=1$ for
axis-parallel spans and $0$ for isotropic ones), and the junction angle between stem and
host. Network statistics are computed at $T=2\times10^4$ (four realisations) unless stated
otherwise; the physical properties at $T=10^3$, with checks at $T=300$ and $3000$.

\subsection{Mechanics and waves}
Each square cell holds $T=10^3$ ligaments, modelled as an Euler--Bernoulli frame with
rigid joints ($E_0=\rho_0=1$); $c_0=\sqrt{E_0/\rho_0}$ is the bar velocity of the
constituent. Spans shorter than their thickness are treated as rigid junction blocks of
length $\lambda$. The boundary of the square is generation zero of the
division, so the material is a periodic tiling of cells whose frames are shared with their
neighbours~\cite{both2026}; we tile by mirror images, which makes the tiling exactly
periodic with period $2L$ and orthotropic. To check that the mirror symmetry does not set
the isotropy, we also compute the full $C$ of the single cell with affine boundary
displacements, which keeps the off-axis terms $C_{16}$ and $C_{26}$. Periodic homogenisation
gives $C$, and the Christoffel equation gives
$n(\hat{\bm k})$, its direction average $\langle n\rangle$ and the isotropy
$\eta_n=\min n/\max n$. Bloch waves along $x$ and $y$ give the band edge $\omega_h$, the
highest frequency at which the longitudinal branch keeps its plane-wave character and its
index within $10\%$ of the static value. The wavenumber grid has 24 points up to
$k_{\max}=\min(2\pi,3n\omega_1)/L$; the median grid step is $15$--$18\%$ of $\omega_h$. The
plane-wave projection unfolds the zone folding of the $2L$ supercell: in $4$--$31$ of the
$60$--$64$ scans of each rule the edge lies above the folded zone-edge frequency $\pi c_0/(2nL)$,
and only $20$ of $247$ edges lie within $10\%$ of it, so the Bragg scattering of the mirror
period does not set $\omega_h$. The clamped cell gives the lowest local resonance
$\omega_1$. Frequencies are in units of $c_0/L$, with $L$ the cell size. We scan
$\alpha\in[0,1]$ and $\bar\rho\in[0.03,0.4]$, four realisations each (eight for the
Bloch scans of SR, \emph{centroid} and \emph{balanced}), and use square,
triangular, honeycomb and Voronoi lattices as references.

\section{Two general laws}
\label{sec:laws}

Two relations hold for every rule and organise the results.

\paragraph*{The index has a ceiling.}
A porous material is at most as stiff as its constituent diluted by the density, the
Voigt bound~\cite{hill1952} $C\le\bar\rho\,C^{0}$, so $n\ge1$ for every frame, with $c_0$ the
bar velocity. There is
no universal upper limit: $n$ diverges as the structure approaches a mechanism. A
longitudinal wave probes $C_{11}\sim K+G$ rather than the Young modulus, so a honeycomb,
soft in $E$ but stiff in the bulk modulus $K$~\cite{deshpande2001}, keeps $n\simeq1.9$ at
every density, whereas a network of T-junctions is soft in both. Its stiffness has two
parts. With pin joints the network carries load by stretching alone, with a longitudinal
modulus $C^{\rm pin}=\bar\rho\,s$ that vanishes if the network is a mechanism for the strain
probed~\cite{maxwell1864}. Rigid joints add the bending of the ligaments, as in stiff fibre
networks~\cite{wilhelm2003,head2003,heussinger2006,broedersz2014}, which grows with their
thickness $\lambda$. Hence
\begin{equation}
n^{-2}=s+c\,(\lambda/\ell)^{q},\qquad n\le n_{\max}=s^{-1/2},
\label{eq:ceiling}
\end{equation}
where $\ell$ is a span length and $c$ and $q$ characterise the architecture. The
pin-jointed reference is the same periodic frame with the bending stiffness multiplied by
$10^{-9}$; $n_{\max}$ changes by less than $0.01$ between factors $10^{-7}$ and
$10^{-11}$. Every junction joins a stem to a host that runs straight through it, so the
pinned frame has mean coordination $z=3$ and, by Maxwell counting, $1/2$ floppy mode per
junction for every rule. What it retains is the stiffness of straight lines that cross the
whole tiling. Under mirror tiling the edges of the square frame become such lines, and in
the \emph{balanced}, \emph{regular} and \emph{rect} divisions so do the first full-width
cuts. The pinned modulus is a fixed number of such lines: $\tfrac12(C_{11}+C_{22})^{\rm
pin}=m\lambda E_0$ with $m=1.5$--$1.9$ for SR, \emph{centroid}, \emph{balanced} and
\emph{regular} at $T=300$, $10^3$ and $3000$ alike. Making the frame $1000$ times thinner
leaves $m=0.01$--$0.14$ for the random-angle rules, $0.4$ for \emph{balanced} and
\emph{regular}, the share of their first cuts, and $3$--$4$ for \emph{rect}. Since
$\lambda=\bar\rho/L_{\rm tot}$ at $\alpha=0$, $s=m/L_{\rm tot}$, and the pinned index
$n_{\max}\propto L_{\rm tot}^{1/2}\propto T^{1/4}$ grows with the number of ligaments: the
direction-averaged pinned index of \emph{balanced} is $5.5$, $7.5$ and $9.8$ at $T=300$,
$10^3$ and $3000$. The ceiling is therefore a property of the tiling and its spanning
lines, not of the division network. Fitted per rule, $q\simeq0.7$--$1.1$, smaller than the
value $2$ of pure bending, so stretching and bending are coupled.
Section~\ref{sec:collapse} identifies the span length $\ell$ that makes the bending term,
$c$ and $q$, the same for every rule.

\paragraph*{The weakest span sets the band.}
The lowest local resonance is the flexural mode of the weakest span between junctions.
A span of length $\ell_s$ and thickness $\lambda_s$ has its lowest mode at
\begin{equation}
\omega_s=\beta^2\frac{\lambda_s}{\sqrt{12}\,\ell_s^2}\,c_0 ,
\label{eq:span}
\end{equation}
with $\beta=\pi$ for pinned ends and $\beta=4.730$ for clamped
ends~\cite{rao2018}. The prediction $\omega_{\rm span}=\min_s\omega_s$ with clamped
ends has no free parameter. It tracks $\omega_1$ over three decades, with a correlation
of $0.979$ between $\ln\omega_1$ and $\ln\omega_{\rm span}$ over all 480 cells. The
correlation is not carried by the differences between rules: within each rule and density,
over the hierarchy exponents and realisations, it is $0.88$--$0.99$ (median $0.96$), and at
$\alpha=0$ and fixed density across rules it is $0.96$--$0.97$. The
ratio $\omega_1/\omega_{\rm span}$ between $0.44=(\pi/4.730)^2$, the pinned limit, and
$0.70$: junctions rotate, so the weakest span is partly clamped. The Bloch band edge
follows the same prediction [Fig.~\ref{fig:phys}(c)], with correlations $0.97$, $0.94$
and $0.97$ for SR, \emph{centroid} and \emph{balanced}, but with a rule-dependent prefactor
$\omega_h/\omega_{\rm span}$ (Sec.~\ref{sec:collapse}). At $\alpha=0$ all ligaments have
the same thickness, and the weakest span is the longest one.

\section{Properties shared by all rules}
\label{sec:shared}

\begin{figure}[tb]
\includegraphics[width=\columnwidth]{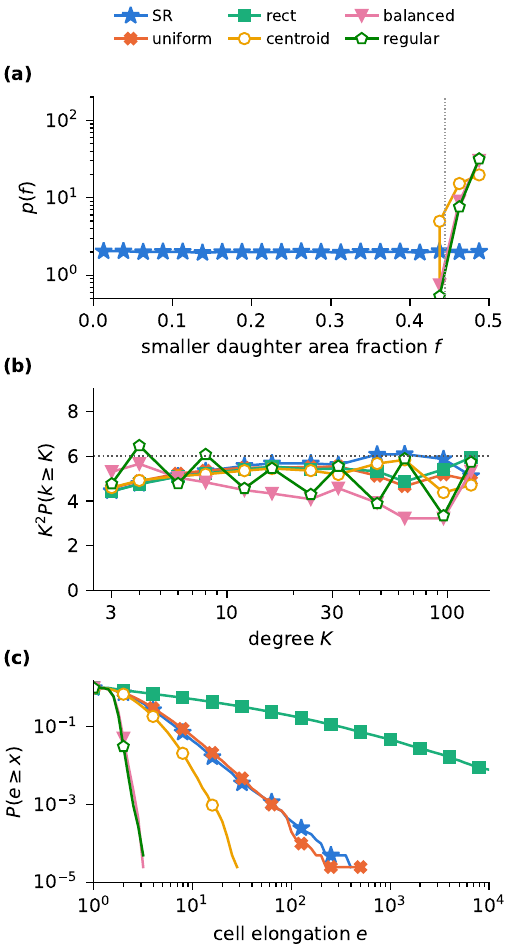}
\caption{\label{fig:net}Network mappings, $T=2\times10^4$, four realisations; symbols as in
Fig.~\ref{fig:rules}. (a) Area fraction $f$ of the
smaller daughter at each split (\emph{uniform} and \emph{rect} coincide with SR); dotted
line: $f=4/9$. (b) Dual-graph degree law, $K^2P(k\ge K)$; dotted line: the asymptote $6$
of $p_k=12/[k(k+1)(k+2)]$. (c) Distribution of the cell elongation $e$ (largest over
smallest width).}
\end{figure}

Several network properties follow from the T-junction geometry alone and are the same for
every rule. They are exact or asymptotically exact, and they were derived for SR in
Ref.~\cite{vazquez2026sr}; we only check that they hold for the other rules.

A junction on the side of a neighbouring cell is a straight angle there, so the corner
number of a cell is fixed at birth, a split gives $n_1+n_2=n+4$, and the mean corner
number is exactly $4$ for every rule. Each ligament has two ends, so the mean dual degree
tends to $4$ (measured $3.91$--$3.95$, the deficit coming from ends on the boundary). A new
ligament ends on two sides of the cell it cuts, and the hit rate of a ligament is
proportional to the number of cell sides it carries, which equals its degree. This gives
the linear attachment kernel $\kappa(k)=a+bk$ with $b=1/2$, and $p_k=12/[k(k+1)(k+2)]$,
$\gamma=3$, for SR~\cite{vazquez2026sr}. What is shared by all rules is the exponent, not
the full distribution. The measured kernel, $t$ times the hit rate of a ligament of degree
$k$ (steps $t>2000$, $2\le k\le20$), is $a+bk$ with $b=0.48$, $0.48$, $0.50$, $0.49$ and
$0.47$ for SR, \emph{uniform}, \emph{rect}, \emph{centroid} and \emph{balanced}, and
$|a|\le0.16$; for \emph{balanced} it is flat up to $k=4$, because a
cut across the long axis avoids the short sides. \emph{regular} cuts its cells generation by
generation, and its kernel is not stationary in time. In every case
$K^2P(k\ge K)$ is flat between $K=8$ and $128$ [Fig.~\ref{fig:net}(b)], so $\gamma=3$, with
an amplitude of $3.2$--$6.1$ against $6$ for the SR law; \emph{balanced} lies lowest, near
$4$. The birth index $h$ of the ligament hit by
the new ligament $t$ follows the law $h/t\to U^2$, with $U$ uniform, for SR; the moments
$\langle(h/t)^a\rangle$ for $a=1/4,1/2,1$ are within $0.06$ of $2/3$, $1/2$ and
$1/3$ for every rule. Mean fewest-turn distances are $7.1$--$8.3$ for all rules.

Since these properties are shared, they cannot explain why the physical properties
differ. In the following sections we look at what does differ: the cell tree, the cell
shapes and the frame graph.

\begin{table}[tb]
\caption{\label{tab:net}Network mappings that distinguish the rules ($T=2\times10^4$; $\ell_*$
at $T=10^3$). $\langle f\rangle$: mean smaller-daughter fraction; $\mathrm{Var}\,d$: variance
of the leaf depth; $C$: dual clustering (corner cuts); $e_{95}$: 95th percentile of the cell
elongation; $a_4$: orientation fabric; $\ell_*\sqrt T$: long-span moment of
Eq.~(\ref{eq:lstar}). Further statistics are given in the text.}
\begin{ruledtabular}
\begin{tabular}{lcccccc}
rule & $\langle f\rangle$ & $\mathrm{Var}\,d$ & $C$ & $e_{95}$ & $a_4$ & $\ell_*\sqrt T$\\
\hline
SR       & 0.25 & 15  & 0.38 & 9.4 & 0    & 3.9\\
uniform  & 0.25 & 16  & 0.45 & 11  & 0.03 & 13\\
rect     & 0.25 & 16  & 0    & 918 & 1    & 18\\
centroid & 0.47 & 0.9 & 0.28 & 6.2 & 0    & 3.2\\
balanced & 0.48 & 0.9 & 0.06 & 2.0 & 0.36 & 1.7\\
regular  & 0.49 & 0.2 & 0.05 & 1.9 & 0.48 & 1.3\\
\end{tabular}
\end{ruledtabular}
\end{table}

\begin{table}[tb]
\caption{\label{tab:phys}Physical properties, $T=10^3$, $\alpha=0$, $\bar\rho=0.2$:
direction-averaged index, isotropy of the single cell, lowest local resonance, and Bloch
band edge at fixed index. The spread of $\langle n\rangle$ between realisations is
$2$--$7\%$, the standard errors of $\omega_h$ are $5$--$16\%$; \emph{rect} is given as a range.
\emph{uniform}: frame model not reliable, Bloch scans not computed; \emph{rect}: realisations
do not overlap in $n$.}
\begin{ruledtabular}
\begin{tabular}{lccccc}
rule & $\langle n\rangle$ & $\eta_n$ & $\omega_1$ & $\omega_h(n{=}3)$ & $\omega_h(n{=}4)$\\
\hline
SR       & 4.4     & 0.56 & 0.44  & 0.37 & 0.20\\
uniform  & 3.4     & 0.62 & 0.20  & --   & --\\
rect     & 4.1--7.9 & 0.29 & 0.007 & --   & --\\
centroid & 4.3     & 0.55 & 0.52  & 0.65 & 0.30\\
balanced & 3.0     & 0.80 & 1.14  & 0.62 & 0.23\\
regular  & 2.7     & 0.77 & 1.23  & 0.42 & 0.16\\
\end{tabular}
\end{ruledtabular}
\end{table}

\section{The six structures}
\label{sec:rules}

\begin{figure}[tb]
\includegraphics[width=\columnwidth]{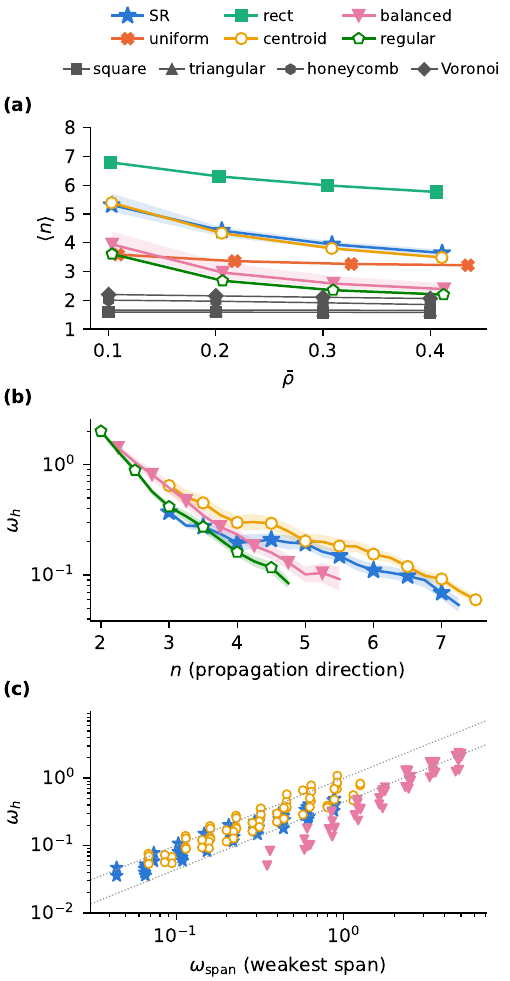}
\caption{\label{fig:phys}Physical properties, periodic tilings, $T=10^3$, $\alpha=0$; symbols as
in Fig.~\ref{fig:rules}, grey: reference lattices. (a) Direction-averaged index against density (frame
included). Lines: geometric mean over four realisations; bands: their range (SR,
\emph{balanced}). (b) Bloch band edge $\omega_h$ (units $c_0/L$) at fixed index along the
propagation direction: each density scan (eight realisations for SR, \emph{centroid} and
\emph{balanced}, four for \emph{regular}; $x$ and $y$) is interpolated
at fixed $n$; lines: mean of $\ln\omega_h$, bands: $\pm$ one standard error. \emph{rect} is
omitted: its realisations do not overlap in $n$. (c)
$\omega_h$ against the weakest-span prediction $\omega_{\rm span}$ of
Eq.~(\ref{eq:span}), every cell; dotted lines: clamped ($\omega_h=\omega_{\rm span}$) and
pinned ($\omega_h=(\pi/4.730)^2\omega_{\rm span}$) spans.}
\end{figure}

For each rule we give the network mapping (Table~\ref{tab:net}, Fig.~\ref{fig:net}), the
physical properties (Table~\ref{tab:phys}, Fig.~\ref{fig:phys}) and the link between them.

\subsection{Scale-rich (SR)}
\emph{Network.} The SR network has been characterised in
Refs.~\cite{both2026,vazquez2026sr}. Its cell tree is a random binary search tree: the
daughter fraction is uniform for any convex cell, and the rescaled areas follow the
spacing law $P(TA>x)=e^{-x}$ of uniform points~\cite{cowan2010,vazquez2026sr}. Corner
cuts make $30\%$ of the cells triangles and give the dual graph its clustering, $C=0.38$.
A cut through a uniform point near a vertex leaves a sliver, so the elongation has a broad
tail, with $0.1\%$ of the cells longer than $66$ times their width, and this tail grows
with $T$ ($e_{99.9}=35$ at $T=10^3$). The longest span, in units of the mean cell size
$T^{-1/2}$, grows from $5.8$ at $T=10^3$ to $8.1$ at $T=2\times10^4$.

\emph{Physics.} The SR cells span $1.70\le n_x\le6.76$ over the scan (with a systematic
uncertainty of about $\pm10\%$ from the beam model, Sec.~\ref{sec:balanced}), consistent
with the reported $1.5$--$8.3$~\cite{both2026} [Fig.~\ref{fig:phys}(a)]. At $\alpha=0$ the
direction-averaged index falls from $\langle n\rangle=5.3$ at $\bar\rho=0.1$ to $3.7$ at
$\bar\rho=0.4$. The hierarchy exponent matters less: at $\bar\rho=0.2$,
$\alpha=0,0.25,0.5,0.75,1$ gives $n_x=4.04,2.84,2.31,2.24,2.36$. The isotropy is
$\eta_n=0.67$ at $\bar\rho=0.2$ for the mirror tiling and $0.56$ for the single cell with
its off-axis terms, which reach $0.30$ of $\sqrt{C_{11}C_{22}}$. Below $n\simeq4.5$ the band
edge at fixed index is the lowest of the random-angle rules [Fig.~\ref{fig:phys}(b)].

\emph{Link.} The long-span tail produced by slivers gives SR its high index at a given
density (Sec.~\ref{sec:collapse}) and its low band edge through Eq.~(\ref{eq:span}).

\subsection{Uniform cell selection}
\emph{Network.} Picking a uniform cell instead of a point keeps the split law and the
tree shape of SR, a random binary search tree, but the areas are no longer spacings:
each cell carries the product of the uniform fractions along its path, so $\ln A$ is a
sum of about $2\ln T$ independent terms. Its variance grows as $4\ln T$ ($27.6$ at
$T=10^3$, measured $23.6$), and the areas spread over many decades. This is the most
scale-rich member of the family. Its dual graph is the same as that of SR
(Fig.~\ref{fig:net}), with a higher clustering, $C=0.45$. At $\bar\rho=0.2$, $89\%$ of the spans
are shorter than twice the ligament thickness and hold $30\%$ of the material: dense
clusters of small cells sit next to large empty cells.

\emph{Physics.} The index is lower than for SR, $\langle n\rangle=3.37$ at $\bar\rho=0.2$,
and nearly independent of density ($3.6$ to $3.2$ from $\bar\rho=0.1$ to $0.4$), and the
resonance is the lowest of the random-angle rules, $\omega_1=0.20$. The frame model is
not reliable for this rule: most spans are stubbier than they are thick, and the clusters
behave as solid inclusions. These numbers are therefore only indicative, and we draw no
conclusion about the physics of this rule beyond its network.

\emph{Link.} The large empty cells give the longest spans of the family after
\emph{rect} ($\ell_{\max}\sqrt T=41$). A continuum model of the clusters would be needed to
say how they affect the index.

\subsection{Axis-parallel cuts (\emph{rect})}
\emph{Network.} With axis-parallel cuts every cell is a rectangle: there are no
triangles, no corner cuts and no dual clustering, $C=0$, and every junction is a right
angle. The split law and the tree are those of SR, but the cell shape is not controlled:
each cut divides either the width or the height by a uniform fraction, so the logarithm
of the elongation performs a random walk and its distribution has no scale [Fig.~\ref{fig:net}(c)], with median $11$ and a tail beyond
$10^4$. This is the multiscaling of the side lengths found by Krapivsky and
Ben-Naim~\cite{krapivsky1994}. Long thin cells are bounded by long spans:
$\ell_{\max}\sqrt T=127$. The orthogonal frame also offers many fewest-turn paths,
$\langle\ln\mu\rangle=3.55$.

\emph{Physics.} The \emph{rect} cells have the highest index, $\langle n\rangle=6.49$ at
$\bar\rho=0.2$, the lowest isotropy, $\eta_n=0.41$, and a resonance two orders of
magnitude below the others, $\omega_1=0.0071$. The index depends strongly on the
realisation and the direction: along $x$ and $y$ the four realisations give $n$ between
$1.7$ and $12.8$ at the same densities, and Table~\ref{tab:phys} gives a
range. The Bloch band edge lies below $0.1$ wherever it is resolved; in half of the scans the
branch keeps its plane-wave character up to the end of the grid, $3\omega_1$, so only that
lower bound is known.

\emph{Link.} The unbounded elongation gives the longest spans and therefore both the
high index and the low resonance (Sec.~\ref{sec:collapse}). The anisotropy follows from the
fabric: for an affine network of the same spans, the stretching modulus
$\sum\ell\cos^4(\phi-\theta)$ gives an isotropy $\eta_{\rm aff}=0.60$, the lowest of the
family.

\subsection{Centroid cuts}
\emph{Network.} A line through the centroid of a convex body in the plane leaves at least
$4/9$ of the area on each side~\cite{grunbaum1960}. The \emph{centroid} split fraction is
therefore confined to $[4/9,1/2]$ [Fig.~\ref{fig:net}(a)]; the minimum over $2\times10^4$
splits is $0.4444$, attained when a triangle is cut parallel to a side. With fractions
close to $1/2$ the area hierarchy disappears: the tree is nearly complete, with leaf-depth
variance $0.92$ against $15.4$ for SR, and the area spread is that of a few
generations, $\mathrm{CV}_A=0.67$. The cut angle is still random, so corner cuts remain
($20\%$ triangles, $C=0.28$) and the dual graph is close to that of SR. Slivers are gone:
the elongation tail is bounded, $e_{99.9}=16$ at both $T=10^3$ and $2\times10^4$, and the
longest span is $6.6\,T^{-1/2}$.

\emph{Physics.} The index coincides with that of SR within $4\%$ at every density
[Fig.~\ref{fig:phys}(a)], and the isotropy is similar, $\eta_n=0.69$ for the mirror tiling
and $0.55$ for the single cell. The band edge at fixed index is higher:
$\omega_h=0.65$, $0.45$, $0.30$ and $0.18$ at $n=3$, $3.5$, $4$ and $5.5$, against $0.37$,
$0.27$, $0.20$ and $0.15$ for SR, with standard errors of $5$--$16\%$. The prefactor $\omega_h/\omega_{\rm span}$ is $0.79$, against $0.63$ for SR.

\emph{Link.} Removing the area hierarchy does not change the index: SR and
\emph{centroid} have the same cut angles, similar long-span tails ($\ell_*\sqrt T=3.9$ and
$3.2$ at $T=10^3$) and the same index. The scale-rich area hierarchy of SR is therefore
not what gives it its index. What the centroid cut removes are the slivers and the
slender spans next to them, which raises the band.

\subsection{Balanced divisions}
\label{sec:balanced}
\emph{Network.} Cutting through the centroid perpendicular to the major axis keeps the
split fraction close to $1/2$ ($\langle f\rangle=0.484$) and, in addition, controls the
shape. A rectangle of aspect ratio $a\ge1$ cut across its long side gives two rectangles
of aspect ratio $\max(a/2,2/a)$. This map sends $[1,2]$ onto itself, alternating between
$[1,\sqrt2]$ and $[\sqrt2,2]$ about the fixed point $\sqrt2$, and brings any $a>2$ into
$[1,2]$: cells stay compact, with $e<2$ for $95\%$ of them and $e\le3.3$ for all. Corner
cuts are rare ($5\%$ triangles, $C=0.06$), junctions are close to right angles
($\langle\sin\psi\rangle=0.96$), and the spans are partly aligned with the square
($a_4=0.36$). This alignment is inherited from the square and is diluted as $T$ grows:
$a_4=0.47$ at $T=10^3$ and $0.36$ at $2\times10^4$ (\emph{regular}: $0.66$ and $0.55$). The
longest span is $3.1\,T^{-1/2}$ and does not grow with $T$.

\emph{Physics.} The \emph{balanced} cells have a low index, $\langle n\rangle=2.97$ at
$\bar\rho=0.2$ and $2.39$ at $\bar\rho=0.4$, against $4.43$ and $3.65$ for SR; only the
balanced and regular divisions reach $n<2.7$. The beam model does not set this ranking.
Timoshenko shear raises $n$ by $2$--$6\%$, and rigid junction zones, modelled by a flexible
length $\ell-\lambda$, lower it by $10$--$12\%$ at $\bar\rho=0.2$--$0.4$; the ratio of the
\emph{balanced} to the SR index stays at $0.64$--$0.68$ in every variant. Of the six rules
they are the most isotropic, $\eta_n=0.78$ for the mirror tiling and $0.80$ for the single
cell, whose off-axis terms are only $0.02$ of $\sqrt{C_{11}C_{22}}$, and they have the highest
resonances, $\omega_1=1.14$. Their pin-jointed ceiling is close to that of SR, but the
bending coefficient $c$ of Eq.~(\ref{eq:ceiling}) is six times that of SR at $\bar\rho=0.4$.
At $n=3$ the band edge matches that of \emph{centroid}, $\omega_h=0.62$ against $0.65$;
above $n\simeq3.5$ it falls faster, $0.35$, $0.23$ and $0.10$ at $n=3.5$, $4$ and $5$. The prefactor $\omega_h/\omega_{\rm span}=0.31$ lies below the pinned
limit.

\emph{Link.} Compact cells give short spans, hence a stiff frame, a low index and high
resonances. Reaching a high index then requires slender ligaments close to the ceiling,
where the band collapses. The uniform span lengths also set the low prefactor: many spans
of nearly the same length resonate together and form a band whose lower edge lies below
the frequency of a single span.

\subsection{Regular divisions}
\emph{Network.} Cutting the largest cell first makes the division deterministic up to the
small angular noise: the tree is nearly complete (leaf-depth variance $0.23$), the areas
nearly equal ($\mathrm{CV}_A=0.25$), and $92\%$ of the cells are quadrilaterals, $97\%$ with
$e<2$. The cell shapes, triangles and clustering are those of
\emph{balanced}; the fabric is more aligned ($a_4=0.48$), and the longest span is the
shortest of the family, $2.2\,T^{-1/2}$.

\emph{Physics.} The \emph{regular} cells have the lowest index of the family,
$\langle n\rangle=2.68$ at $\bar\rho=0.2$, isotropy $\eta_n=0.75$, and the highest
resonance, $\omega_1=1.23$. At fixed index, however, its band edge lies below that of
\emph{balanced}: $\omega_h=0.89$, $0.42$, $0.27$ and $0.16$ at $n=2.5$, $3$, $3.5$ and $4$,
against $1.05$, $0.62$, $0.35$ and $0.23$.

\emph{Link.} Removing the remaining area spread of \emph{balanced} shortens the longest
span further, which lowers the index by another $10\%$ and raises the resonance by $8\%$
at fixed density. At fixed index the more uniform spans resonate together even more
strongly than in \emph{balanced}, and the band edge falls: the most uniform division is not
the best one.

\section{From the frame graph to the physics}
\label{sec:collapse}

\begin{figure}[tb]
\includegraphics[width=\columnwidth]{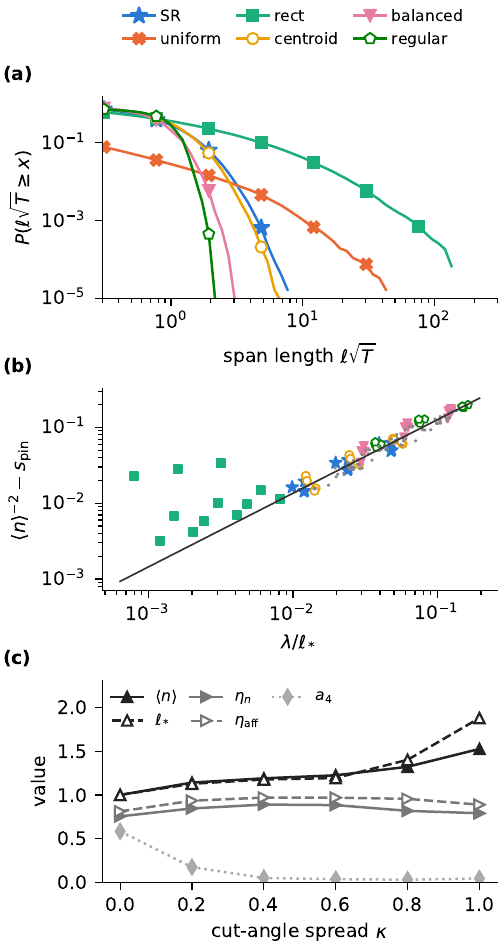}
\caption{\label{fig:link}From the network to the physics. (a) Distribution of span lengths
in units of the mean cell size, $T=2\times10^4$; symbols as in Fig.~\ref{fig:rules}. (b)
Index of every cell ($T=10^3$, $\alpha=0$, $\bar\rho=0.1$--$0.4$, four realisations)
against the slenderness of its long spans, $\lambda/\ell_*$ with $\ell_*$ from
Eq.~(\ref{eq:lstar}); the stretching term $s_{\rm pin}$ of each cell is subtracted, leaving
the bending term. Large symbols: the five frame rules; small grey dots: the cut-angle
family; line: $c\,(\lambda/\ell_*)^q$ fitted to SR, \emph{centroid}, \emph{balanced},
\emph{regular} and the cut-angle family (\emph{rect} shown, not fitted). (c) The cut-angle family at $\bar\rho=0.2$: index and
$\ell_*$ relative to $\kappa=0$, isotropy $\eta_n$, affine isotropy $\eta_{\rm aff}$ of the
span fabric, and fabric $a_4$.}
\end{figure}

\subsection{The long spans set the index}
The previous section suggests a single cause for the differences: the tail of long spans
in the frame graph [Fig.~\ref{fig:link}(a)]. Bending compliance grows as $\ell^3$, so the
longest spans dominate the softness of the frame. We take as length scale in
Eq.~(\ref{eq:ceiling}) a high moment of the length-weighted span distribution,
\begin{equation}
\ell_*=\left(\frac{\sum_s\ell_s^{p+1}}{\sum_s\ell_s}\right)^{1/p},\qquad p=10,
\label{eq:lstar}
\end{equation}
which approaches the longest span as $p$ grows. We fit the rules whose long-span tail is
bounded, that is, whose elongation and $\ell_*\sqrt T$ stay finite as $T$ grows: SR,
\emph{centroid}, \emph{balanced}, \emph{regular} and the cut-angle family. We fix $s$ at
the pinned value $s_{\rm pin}$ of each cell (Sec.~\ref{sec:laws}), so that only the bending
term is fitted. With $\ell=\ell_*$ it is a single power law,
$n^{-2}-s_{\rm pin}=c\,(\lambda/\ell_*)^q$ with $c=1.18$ and $q=0.97$, for every rule and
density [Fig.~\ref{fig:link}(b)]. The root-mean-square deviation in $\ln n$ is $0.059$,
against $0.047$ for separate three-parameter fits to each rule and a total spread of
$0.26$, and the mean offset of each rule is at most $0.055$. A common fitted $s$ instead of
$s_{\rm pin}$ gives the same $c$, $q$ and rms, because at $T=10^3$ the rules have similar
$L_{\rm tot}$; the bending term does not depend on how the ceiling is handled. The law
also predicts rules it has not seen: fitted without one rule, it predicts that rule with
rms $0.057$--$0.084$ in $\ln n$. The residuals of each rule depend little on density
(mean offsets $\le0.09$ at every density). The choice of $\ell$ is what makes the law
universal. With the mean span instead of $\ell_*$ the rms is $0.131$, and with $\lambda$
alone as the abscissa, no network length at all, it is $0.152$. The rms falls with $p$ and
saturates for $p\ge6$ ($0.106$, $0.094$, $0.076$, $0.066$, $0.060$ and $0.061$ for
$p=1$, $2$, $4$, $6$, $10$ and $20$; $0.064$ for the longest span). The rule therefore
enters the index through $\ell_*$: the network property that distinguishes the index of
the divisions is the long-span tail of the frame graph. How the disorder of the segments changes the stiffness of
bending-dominated structures has been studied for Voronoi honeycombs, where a
non-periodic arrangement of the walls leaves the moduli unchanged on
average~\cite{silva1995}, for Voronoi foams, where irregular cells raise the Young and shear
moduli at fixed density and lower the bulk modulus~\cite{zhu2000}, and for fibre networks, where nonaffine floppy-mode deformations bend the segments and the
modulus follows from the segment-length distribution~\cite{heussinger2006,broedersz2014}. What is new
here is that one moment of the span distribution places divisions generated by different
rules on one curve, and that the division rule sets this moment through the cell shapes.

Two rules are outside the fit. The \emph{uniform} rule is left out because its frame model
fails (Sec.~\ref{sec:rules}). The long-span tail of \emph{rect} is unbounded, its index varies
by up to $7\times$ between directions, and a direction average is a poor summary. Its cells
stay below their own pinned ceilings and follow the trend of the law
[Fig.~\ref{fig:link}(b)], but they scatter by a factor of about $4$ in the bending term.

The law also follows the number of ligaments per cell. At fixed density the index rises with
$T$ for every rule: at $\bar\rho=0.2$, $T=300$, $10^3$ and $3000$ give $3.80$, $4.43$ and
$4.80$ for SR, $3.94$, $4.34$ and $4.73$ for \emph{centroid}, $2.68$, $2.97$ and $3.10$ for
\emph{balanced} and $2.43$, $2.68$ and $2.84$ for \emph{regular}. The ranking of the rules
is the same at every $T$. Part of this drift is the frame ceiling, which rises as $T^{1/4}$.
With the pinned ceiling of each cell, the bending law fitted at $T=10^3$ predicts the cells
at $T=3000$ with rms $0.04$--$0.07$ in $\ln n$ and at $T=300$ with rms $0.05$--$0.12$,
without refitting; a common fixed $s$ does worse at $T=300$ ($0.11$--$0.18$). The index of a
rule is therefore not a single number. The bending term is fixed by $\lambda/\ell_*$, which
depends on $\bar\rho$, on $T$ and, for SR and \emph{rect}, on the growth of the long-span
tail with $T$, and the ceiling by the frame of the tiling. The bending collapse does not
need the frame: with a frame $1000$ times thinner, the index follows a pure power law,
$n^{-2}=1.6\,(\lambda/\ell_*)^{1.11}$, with rms $0.11$ and rule offsets $\le0.10$.

Within one rule the realisations differ less, and the total ligament length matters
more. At $\bar\rho=0.2$ and $\alpha=0$ the thickness is $\lambda=\bar\rho/L_{\rm tot}$, so
a realisation with more ligament length has slenderer spans: over 16 realisations of
\emph{balanced} and \emph{regular}, the correlation between $\ln n$ and the mean span
length, proportional to $L_{\rm tot}$, is $0.94$ and $0.97$. For SR and \emph{centroid} the
spread between realisations, $2$--$4\%$ in $n$, is too small to test.

\subsection{The same spans set the band}
At $\alpha=0$ the weakest span of Eq.~(\ref{eq:span}) is the longest span, so $\ell_*$
also sets the band edge, $\omega_h\propto\lambda/\ell_*^2$. Eliminating $\lambda$ with
Eq.~(\ref{eq:ceiling}) gives the index--bandwidth trade-off
\begin{equation}
\omega_h(n)\simeq\frac{\Omega}{\ell_*}\,\bigl(n^{-2}-n_{\max}^{-2}\bigr)^{1/q},
\label{eq:tradeoff}
\end{equation}
with $\Omega$ the prefactor of Eq.~(\ref{eq:span}) times $c^{-1/q}$. At equal index, a
rule with shorter long spans has the higher resonance, provided it reaches that index far
from its ceiling. Fitted to the clamped resonances, Eq.~(\ref{eq:tradeoff}) describes the
\emph{balanced} and \emph{regular} divisions within $1$--$9\%$, and the scale-rich
divisions within about $30\%$, where the weakest span varies from realisation to
realisation. For the Bloch edge, the prefactor $\omega_h/\omega_{\rm span}$ differs between
rules by a factor of $2.5$: it is $0.79$ for \emph{centroid}, $0.63$ for SR and $0.31$ for
\emph{balanced}. We attribute this to how many spans share the weakest frequency, but we do
not predict it. Equation~(\ref{eq:tradeoff}) therefore organises the trade-off; the ranking
of the rules at equal index in Sec.~\ref{sec:best} is measured, not derived.

\subsection{The cut angle as a continuous variable}
The cut-angle family keeps the centroid cut and the area tree fixed ($\langle f\rangle$
between $0.473$ and $0.486$, leaf-depth variance $0.87$--$0.93$) and changes only the
angle. As $\kappa$ grows from $0$ to $1$, triangles become more frequent ($C$ rises from
$0.05$ to $0.28$), the junctions become oblique ($\langle\sin\psi\rangle$ falls from $0.96$
to $0.80$), the elongation tail grows ($e_{95}$ from $2.0$ to $6.0$), and $\ell_*$ grows by
$1.9\times$ [Fig.~\ref{fig:link}(c)]. The index rises monotonically, from $2.94$ to $4.41$ at
$\bar\rho=0.2$, from $3.98$ to $5.45$ at $\bar\rho=0.1$ and from $2.37$ to $3.57$ at
$\bar\rho=0.4$, with a spread of $0.01$--$0.15$ between realisations: within one family,
$2.4\le n\le5.5$. The Bloch band edge at $\bar\rho=0.2$ falls from $1.00$ to $0.29$. Each
of these network quantities correlates with $n$ across the family (correlations of
magnitude $0.86$--$0.98$), but only $\ell_*$ also collapses the other rules: the clustering,
for example, vanishes for \emph{rect}, which has the highest index.

The isotropy is not monotone: $\eta_n=0.76$ at $\kappa=0$, $0.89$ at $\kappa=0.4$ and
$0.79$ at $\kappa=1$. Two network properties explain the maximum. The fabric of the
perpendicular cuts is aligned with the square ($a_4=0.58$ at $\kappa=0$), and a small
angular spread suffices to make it isotropic ($a_4=0.05$ at $\kappa=0.4$); the affine
isotropy of the fabric follows, $\eta_{\rm aff}=0.81$, $0.97$ and $0.89$. Beyond
$\kappa\simeq0.6$ the fabric stays isotropic but the elongation tail and $\ell_*$ grow, and
the long spans, few and differently oriented in each realisation, make the cell
anisotropic. The most isotropic cells combine an isotropic fabric with compact cells. The
maximum is not produced by the mirror symmetry of the tiling. For the single cell with its
off-axis terms, whose $\langle n\rangle$ agrees with the periodic value within $4\%$,
$\eta_n=0.77$, $0.90$, $0.90$, $0.88$, $0.75$ and $0.52$ for $\kappa=0$, $0.2$, \dots, $1$,
a plateau between $\kappa=0.2$ and $0.6$; the off-axis terms grow from $0.02$ to $0.39$ of
$\sqrt{C_{11}C_{22}}$, so the mirror tiling overstates the isotropy of the random-angle cells
and makes the maximum less pronounced, not more.

\section{Which structure is best}
\label{sec:best}

The ranking at equal index is measured [Fig.~\ref{fig:phys}(b)]; the network properties
that set $\ell_*$ explain its main features. Below $n\simeq2.7$ only the
\emph{balanced} and \emph{regular} divisions are available, with $\omega_h$ up to about $2$;
\emph{balanced} keeps the wider band, $1.05$ against $0.89$ at $n=2.5$ (standard errors
$6$--$9\%$). At $n=3$ \emph{balanced} and \emph{centroid} tie ($0.62$ and $0.65$), and SR
($0.37$) and \emph{regular} ($0.42$) trail. From $n\simeq3.5$ upwards \emph{centroid} has the
highest band edge, $0.45$, $0.30$, $0.29$ and $0.20$ at $n=3.5$, $4$, $4.5$ and $5$, against
$0.35$, $0.23$, $0.16$ and $0.10$ for \emph{balanced}, with standard errors of
$11$--$16\%$: the advantage is about $1.5$ standard errors at $n=3.5$--$4$ and larger above.
SR ties \emph{balanced} at $n=4$, exceeds it from $n\simeq4.5$, and is within one standard
error of \emph{centroid} at $n=5$.
The random cut angle provides the index through elongated cells and long spans; the
centroid cut removes the slivers that make the weakest spans slender. Neither the area
hierarchy nor the extra scales of the \emph{uniform} rule help, and the unbounded
elongation of \emph{rect} gives a high index at the price of the band and of the
isotropy. The reference lattices stay below $n=2.1$, well apart from the divisions even
allowing for the $\pm10\%$ uncertainty of the beam model [Fig.~\ref{fig:phys}(a)]: square
$1.41$, triangular $1.72$, honeycomb $1.9$--$2.1$ and Voronoi $2.1$ along $x$.

\section{Three dimensions}
\label{sec:3d}

\begin{figure}[tb]
\includegraphics[width=\columnwidth]{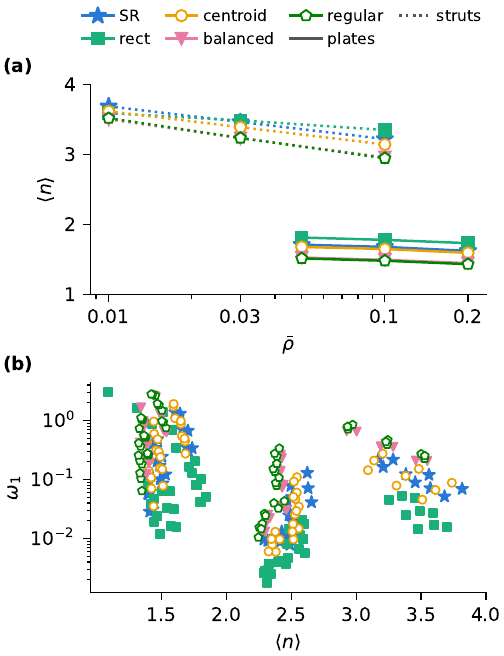}
\caption{\label{fig:3d}Three dimensions, divisions of the cube, $T=300$; symbols as in
Fig.~\ref{fig:rules}. Solid lines: plates; dotted lines: struts; in (b) the plates lie at
$\langle n\rangle<2$ and the struts above. (a) Direction-averaged index against density, $\alpha=0$. (b) Lowest local
resonance $\omega_1$ (units $c_0/L$) against $\langle n\rangle$, $\alpha\in\{0,1/2,1\}$.}
\end{figure}

A recursive division of the cube cuts convex polyhedra by planes, so the material is a
closed-cell network of plates~\cite{both2026}; its edges form an open-cell network of
struts. The dual graph of the plates has a degree exponent $\gamma=5/2$ for SR, and the
host-age and face-local arguments carry over with $b=(d-1)/d$~\cite{vazquez2026sr}. We use
the same rules, with isotropic, axis-parallel or major-axis normals, and $T=300$ plates
per cell. Plates are meshed with triangles that conform across T-junction
lines~\cite{shewchuk1996} and modelled by constant-strain membranes and discrete
hinges~\cite{grinspun2003} ($\nu_0=0.3$); struts are three-dimensional Euler--Bernoulli
beams. The moduli converge to 1--2\% under mesh refinement, change by less than $1\%$
from $T=300$ to $600$, and agree within 1--3\% between affine and periodic boundaries. The
\emph{uniform} rule is excluded: its plates are thicker than they are wide.

In three dimensions the ceiling of Eq.~(\ref{eq:ceiling}) is low (Fig.~\ref{fig:3d}). A
closed convex polyhedron with rigid faces is rigid (Cauchy's theorem~\cite{connelly1993}),
so, as in periodic plate lattices~\cite{berger2017,tancogne2018}, membranes alone carry the
load and the plates are stretching dominated. With bending
switched off, SR plates give $n_{\max}=1.73$ and \emph{balanced} plates $1.58$; the full
model gives $1.71$ and $1.54$ at $\bar\rho=0.05$. Over the scan, $\alpha\in\{0,1/2,1\}$
and $\bar\rho\in\{0.05,0.1,0.2\}$, every rule stays within $1.32\le\langle n\rangle\le1.84$.
The open-cell struts give $2.3\le\langle n\rangle\le3.7$ for every rule, nearly
independent of density, the signature of a stretching-dominated response with a low
ceiling; we have not computed their pinned index. The network then matters through the
spans alone: at equal $n$ the balanced divisions, with their compact cells, have the higher
lowest resonance, $\omega_1=1.30$ against $0.20$ for SR plates at $\langle n\rangle\simeq1.5$
and $0.22$ against $0.06$ for struts at $\langle n\rangle\simeq3.6$. Bloch band edges were
not computed in three dimensions.

\section{Conclusions}

The recursive divisions form a family whose members share their dual-graph degree law,
mean degree, mean corner number and host-age law, and differ in their cell trees, cell
shapes and span distributions. The scale-rich area hierarchy of the SR rule, and the
still broader one of uniform cell selection, are not what sets the physics. The physics
follows the tail of long spans in the frame graph. With these spans as the length scale,
the bending part of the inverse squared index follows one power law for every rule with a
bounded tail, at every density and number of ligaments, and the same spans set the lowest
local resonance. The ceiling of the index is set by the straight frame lines of the tiling,
not by the division. The cut angle controls the tail: random angles elongate the cells and give a high
index and a narrow band, perpendicular cuts keep the cells compact and give a low index
and a wide band, and axis-parallel cuts let the elongation grow without bound. The cut
position controls the slivers: a centroid cut bounds the split fraction between $4/9$ and
$5/9$, keeps the index of random angles and widens the band. Balanced divisions are
therefore best at low index and centroid divisions at high index, and the cut angle
interpolates continuously between them, with the most isotropic cells in between. In
three dimensions the plate and strut divisions are stretching dominated, the ceiling is
low, and the compact cells of balanced divisions keep the higher resonances. Above the band edge,
local resonances open gaps~\cite{liu2000}, a separate design problem in which the broad
span spectrum of a scale-rich division may be an asset.

\begin{acknowledgments}
All code is available at \url{https://github.com/av2atgh/scalerich/tree/master/beyond}.
The calculations were assisted by Claude Opus~5.5, an AI assistant developed by Anthropic.
Nodes \& Links Ltd provided support in the form of salary for Alexei Vazquez but did not
have any additional role in the conceptualization of the study, the analysis, the decision
to publish or the preparation of the manuscript.
\end{acknowledgments}

\bibliography{refs}

\end{document}